\documentclass[a4paper,fleqn,usenatbib]{mnras}

\usepackage{newtxtext,newtxmath}

\usepackage[T1]{fontenc}
\usepackage{ae,aecompl}
\usepackage{color}

\usepackage{graphicx}	
\usepackage{amsmath}	
\usepackage{amssymb}	
\usepackage{subfigure}
\usepackage{natbib}
\usepackage{booktabs}

\def\aj{AJ}%
\def\apj{ApJ}%
\def\apjl{ApJ}%
\def\apjs{ApJS}%
\def\ao{Appl.~Opt.}%
\def\aap{A\&A}%
\def\mnras{MNRAS}%
\def\pasp{PASP}%
\def\nat{Nature}%
\def\procspie{Proc.~SPIE}%

\defcitealias{Carnall2015}{C15}
\newcommand{\angstrom}{\textup{\AA}}

\title[ ATLAS $+$ WISE $z>6$ quasars]{Two more, bright, $z>6$ quasars from VST ATLAS and WISE}

\author[B. Chehade et al.]{
B. Chehade,$^{1}$\thanks{E-mail: benchehade@gmail.com}
A. C. Carnall,$^{1, 2}$\thanks{E-mail: adamc@roe.ac.uk}
T. Shanks,$^{1}$\thanks{E-mail: tom.shanks@durham.ac.uk}
C. Diener,$^3$
M. Fumagalli,$^{1, 4}$
\newauthor
J.R. Findlay,$^5$
N. Metcalfe,$^1$
J. Hennawi,$^6$
C. Leibler,$^7$
D.N.A. Murphy,$^3$
\newauthor
J.X. Prochaska,$^7$
M.J. Irwin$^3$
\& E. Gonzalez-Solares$^3$
\\
$^1$Centre for Extragalactic Astronomy, Department of Physics, Durham University, South Road, Durham DH1 3LE, UK\\
$^2$Institute for Astronomy, University of Edinburgh, Royal Observatory Edinburgh, UK\\
$3$ Institute of Astronomy, Madingley Road, Cambridge CB3 0HA, UK\\
$^4$Institute for Computational Cosmology, Department of Physics, Durham University, South Road, Durham DH1 3LE, UK\\
$^5$Department of Physics and Astronomy, University of Wyoming, 1000 E. University, Department 3905, Laramie, WY 82071, USA\\
$^6$Max-Planck-Institut fur Astronomie, Konigstuhl 17, D-69117 Heidelberg, Germany\\	
$^7$University of California Observatories-Lick Observatory, University of California, 1156 High Street, Santa Cruz, CA 95064, USA; USA\\
}
\date{Accepted 2018 February 23. Received 2018 February 22; in original form 2017 June 1.}

\pubyear{2018}

\begin{document}
\label{firstpage}
\pagerange{\pageref{firstpage}--\pageref{lastpage}}
\maketitle

\begin{abstract}
Recently, Carnall et al. discovered two bright high redshift quasars
using the combination of the VST ATLAS and WISE surveys. The technique
involved using the 3-D colour plane $i-z:z-W1:W1-W2$ with the WISE
$W1$(3.4 micron) and $W2$ (4.5 micron) bands taking the place of the
usual NIR J band to help decrease stellar dwarf contamination. Here we
report on our continued search for $5.7<z<6.4$ quasars over an
$\approx2\times$ larger area of $\approx3577$ deg$^2$ of the Southern
Hemisphere. We have found two further $z>6$ quasars, VST-ATLAS
J158.6938-14.4211 at $z=6.07$ and J332.8017-32.1036 at $z=6.32$ with magnitudes
of $z_{AB}=19.4$ and 19.7 mag respectively. J158.6938-14.4211 was confirmed
by Keck LRIS observations and J332.8017-32.1036 was confirmed by ESO NTT
EFOSC-2 observations. Here we present VLT X-shooter Visible and NIR
spectra for the four ATLAS quasars. We have further independently
rediscovered two $z>5.7$ quasars previously found by the VIKING/KiDS and
PanSTARRS surveys. This means that in ATLAS  we have now discovered a
total of six quasars in our target $5.7<z<6.4$ redshift range.  Making
approximate corrections for incompleteness, we find that our quasar
space density agrees with the SDSS  results of Jiang et al.  at
$M_{1450\angstrom}\approx-27$. Preliminary virial mass estimates based
on the CIV and MgII emission lines give black hole masses in the range
$M_{BH}\approx1-6\times10^9M_\odot$ for the four ATLAS quasars.
\end{abstract}


\begin{keywords}
quasars: general - quasars: individual: VST-ATLAS J158.6938-14.4211  - quasars: individual: VST-ATLAS J332.8017-32.1036
\end{keywords}



\section{Introduction}


High redshift quasars are key probes of the Universe within the first
billion years to redshift $z \sim6$. Firstly, they provide observational
constraints on the evolution of the quasar luminosity and black hole mass
functions to the highest redshifts (\citealt{Willott2005},
\citeyear{Willott2010a}, \citeyear{WIllott2010b}, \citealt{Jiang2009},
\citeyear{Jiang2016}). The existence of a high spatial density of
luminous quasars at $z>6$ would be a challenge to the standard
cosmological model, which is limited in its capacity to produce large
black hole masses, assuming Gaussian initial conditions (e.g.
\citealt{Rosas2016}).

Secondly, analysis of damped Ly$\alpha$ systems (DLAs) and Lyman Limit
Systems (LLS) along quasar lines of sight at intermediate redshifts
($z\approx3$) provide invaluable constraints on galaxy evolution by
probing conditions within the interstellar medium (ISM) and circumgalactic medium (CGM)
(e.g. \cite{Fumagalli2016}, \cite{Wotta2016}, \cite{Lehner2016} and
references therein), via absorber metallicities, covering factors
and kinematics. Some progress has been made in extending these analyses
to higher redshifts (e.g. \cite {Rafelski2014},\cite{Crighton2015},
\cite{Wang2015}).

Finally, high redshift quasars cast light on conditions in the intergalactic
medium (IGM) during the epoch of reionisation. Studies have been made of
the Ly$\alpha$ emission-line profiles of these objects, their ionized
near zones and the Gunn-Peterson troughs in their spectra (e.g.
\cite{Becker2001}, \cite{Fan2006}, \cite{Carilli2010},
\cite{Venemans2015a} and \cite{Barnett2017}) producing a general
consensus view of a patchy reionisation which ended around $z=5-6$. 

There has also been much debate about the relative contributions of
active galactic nucleus (AGN) activity and star formation to the
reionization of the IGM, and the luminosity function of high redshift
quasars is a crucial variable in this context. Whilst it has been shown
(\citealt{McLeod2015}, \citeyear{McLeod2016}) that star formation in
high redshift galaxies can provide the bulk of the ionizing photons
necessary for reionization, the actual fractional contributions from AGN
and star formation are still unknown, with \cite{Madau2015} presenting a
plausuble scenario in which AGN activity provides the majority of the
ionizing photons. The major extra AGN contribution in their work is due
to the increased slope of the faint end of the quasar luminosity
function found in the CANDELS survey by \cite{Giallongo2015}.  However,
using the same CANDELS data, \cite{Parsa2017} report no evidence for a
significant AGN contribution. Although here we shall be addressing the
bright end only, the importance of determining the quasar luminosity
function at high redshift is clear not only for the source  of
reionising photons but also for models of black hole seeding, growth,
feedback and evolution. Clearly, bright, high redshift, quasars are also
easier to follow-up for the above ISM, CGM and IGM absorption
studies.


The techniques employed to detect high redshift quasars are based on
essentially the same Lyman-dropout technique as has been applied for
many years, where objects with a very red colour, often implied by a
non-detection in a bluer band, are selected from large area sky surveys.
One of the first applications of this technique was by \cite{Shanks1983}
who looked for $U$ and $B$ dropouts relative to their $R$ band
magnitudes to find a quasar at $z=3.63$. This method has evolved more
recently to redder dropout bands in order to probe higher redshifts.
However, due to the decreasing spatial density of quasars at
increasingly high redshifts, contaminant objects such as cool L and T
dwarf stars become more and more problematic. Due to this, a variety of
new methods have recently been developed for cleaning photometrically
selected samples before spectroscopic follow-up observations are made to
confirm the nature of the sources.

The first  study to be sensitive to quasars above $z=6$ was reported
by \cite{Fan2001, Fan2006} and \cite{Jiang2016} using data from the Sloan
Digital Sky Survey (SDSS).  They looked for $i$ band dropouts relative
to their $z$ band magnitudes and then obtained follow-up $J$ band
imaging, selecting quasars as objects with blue $z-J$ colour to exclude
cool dwarf stars. The same approach has also been applied to other
survey data e.g. by \cite{Willott2010a}, \cite{Venemans2015b} and
\cite{Wang2017}. More recently, large area $Y$ band surveys have been used
instead of follow-up $J$ band photometry by \cite{Banados2014,Banados2016} 
and \cite{Reed2015}. A Bayesian statistical approach to
candidate selection has also been developed by \cite{Mortlock2012}
and \cite{Matsuoka2016}. Most recently the dropout method has been
applied to the $z$ band by \cite{Venemans2013, Venemans2015a} to begin
selecting quasars at $z\gtrsim6.5$, with the current record holder being
a quasar at $z=7.54$ discovered by \cite{Banados2017}.



In this paper we follow \cite{Carnall2015} (hereafter
\citetalias{Carnall2015}) who found two $z>6$ quasars using the
combination of the new Very Large Telescope Survey Telescope ATLAS
(VST ATLAS, \citealt{Shanks2015}) and Wide-field Infrared Survey
Explorer (WISE, \citealt{Wright2010}) surveys, by applying a similar
method to a greater area of VST ATLAS imaging which yielded the discovery of two more
bright $z>6$ quasars. 

In Section \ref{surveydata} we review the survey data, and in Section
\ref{quasarselection} we review our quasar selection technique. In
Section \ref{spectroscopy} we report on follow-up spectroscopic
observations of two further high redshift quasar candidates, which we
confirm to be two new quasars at $z=6.32$ and $z=6.07$. In Section
\ref{qlf} we make an approximate estimate of the bright end of the
quasar luminosity function from our new quasars plus the others found
within our search area. We present our conclusions in Section
\ref{conclusions}.

All VST ATLAS magnitudes are given on the AB system, all other
magnitudes are on the Vega system. All cosmological calculations assume
the density parameters $\Omega_m=0.3$, $\Omega_\Lambda=0.7$ and Hubble Constant 
$h=0.7$, with $h$ measured in units of 100kms$^{-1}$Mpc$^{-1}$.

\begin{figure}
	\includegraphics[width=\linewidth]{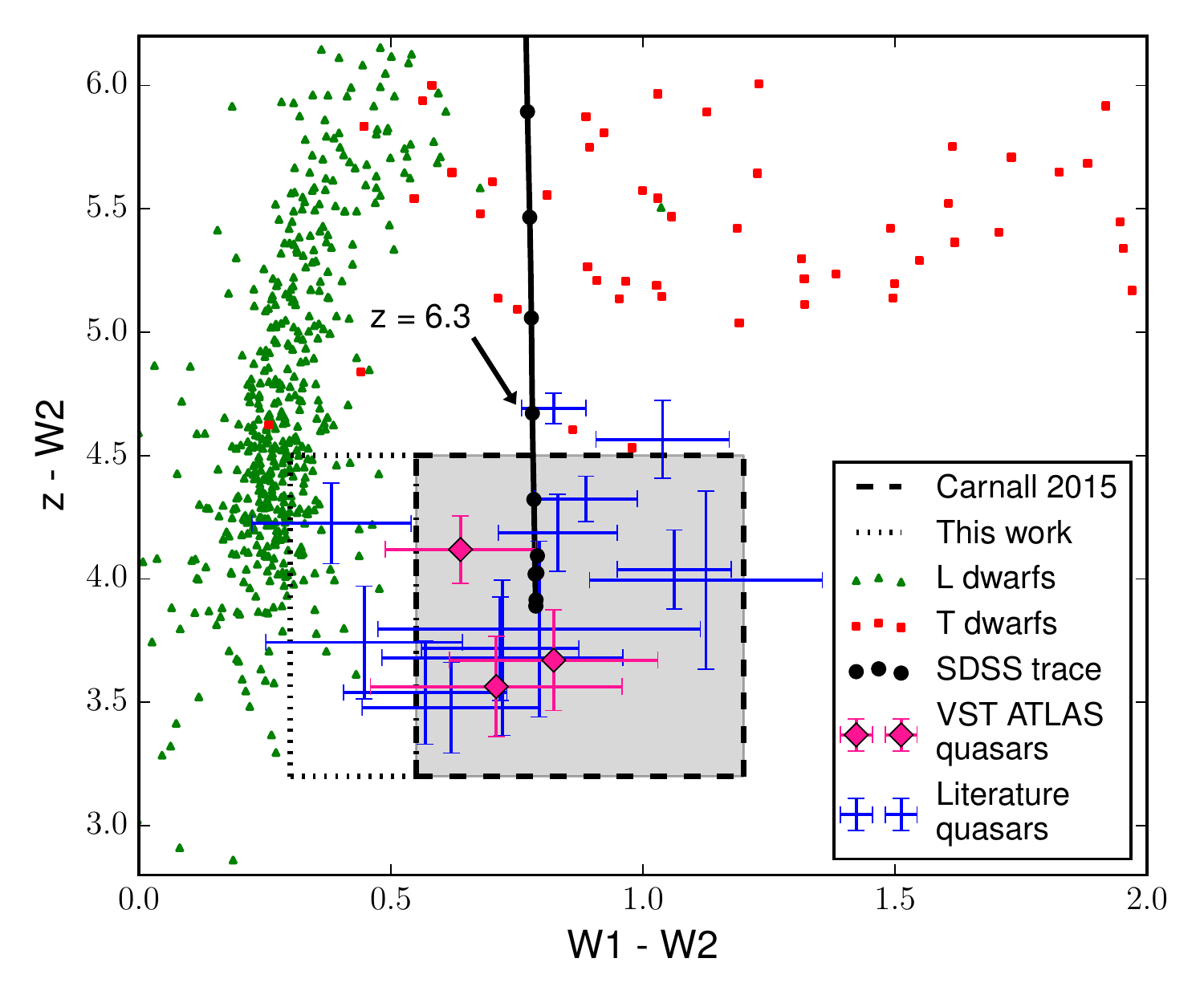}
	\caption{The $z-w2:W1-W2$ colour plane with the positions of  4
	ATLAS quasars marked. Also marked are the positions of L and T dwarf
	stars and other quasars from the literature. The dashed-line box
	shows  the colour selection of \protect\citetalias{Carnall2015}. 
	The dotted line extends this selection bluewards in $W1-W2$ to
	$W1-W2>0.3$, including two previously discovered quasars from SDSS
	and with the aim of testing the completeness of the  \protect\citetalias{Carnall2015}
	selection. The black line is predicted quasar redshift track 
	marked in 0.1 redshift intervals from $z=5.8$ to $z=6.6$.}
	\label{fig:colselection}
\end{figure}

\begin{table*}
\begin{center}
\begin{tabular}{cccccccccc}
\hline
 Quasar & $i$ (AB) & $z$ (AB) &Y&J&H&K & $W1$  & $W2$ \\
\hline
 VST-ATLAS J332.8017-32.1036  & \textgreater21.84  & $19.75\pm0.06$ &$19.43\pm0.06$&$18.98\pm0.05$&$18.43\pm0.07$&$17.85\pm0.07$&$16.36\pm0.05$&$15.57 \pm 0.09$ \\
 VST-ATLAS J158.6938-14.4211  & \textgreater21.56 & $19.44\pm0.08$ &$18.78\pm0.07$&$18.27\pm0.06$&$17.58\pm0.09$&$16.86\pm0.08$&$15.86\pm0.08$ &$15.13\pm0.13$ \\
 PSO J340.2041-18.6621        & \textgreater22.5  & $19.44\pm0.10$ &-             &-             &-              &$17.69\pm0.19$&$16.56\pm0.07$&$15.53\pm0.13$ \\
\hline
\end{tabular}
\end{center}
\caption{VST-ATLAS (i,z), VHS (Y,J,H,K) and ALLWISE (W1,W2) magnitudes for three quasars newly selected from the VST ATLAS survey.
None of the  quasars were detected in the $i$ band and 3$\sigma$ limiting magnitudes are provided. PSO J340.2041-18.6621
was previously discovered by \protect \cite{Banados2014}. All magnitudes are on the Vega system except where indicated.}
\label{table:table1}
\end{table*}

\section{Survey Data}
\label{surveydata}
\subsection{VST ATLAS}
The VLT Survey Telescope (VST) is a 2.6 m wide-field survey telescope
with a $1^{\circ} \times 1^{\circ}$ field of view. The OmegaCAM camera
\citep{Kuijken2004} consists of $32$ CCDs with $2k \times 4k$ pixels,
resulting in $16k \times 16k$ image with a pixel scale of $0.''21$.
The VST ATLAS is a nearly completed photometric survey that will cover
$\approx 4700$ deg$^{2}$ of the southern extragalactic sky with coverage
in $ugriz$ bands. The survey takes two sub-exposures of $2\times60$s
($u$), $2\times50$s ($g$), $2\times45$s ($r$), $2\times45$s ($i$),
$2\times45$s ($z$), per 1 degree field with a small dither in X and Y to
cover most interchip gaps. The sub-exposures are then reduced and
stacked by the Cambridge Astronomy Survey Unit (CASU). Their pipeline
outputs catalogues cut at $\approx5\sigma$ and provide fixed aperture
fluxes and morphological classifications of detected objects (see
\cite{Shanks2015} for more details). Here, for stellar photometry, we
use a $1 ''$ radius aperture (i.e. {\tt aper3}). ATLAS photometry is
calibrated using the APASS Nightly zero-points ({\tt NIGHTZPT} in the
FITS headers, see \citealt{Shanks2015}). The  star-galaxy classification
is that supplied as default in the CASU catalogues and discussed
in detail \citep[by][]{Gonzalez2008}. 

VST ATLAS  provides $ugriz$ photometry to similar depths as SDSS for
galaxies and up to 0.7 mag fainter (e.g. in the $z$-band) for stars,
mainly due to its $\approx40$\% better seeing (e.g. $0.''81$ versus
$1.''18$ in $i$, \citealt{Shanks2015}). In terms of VST${-}$ATLAS progress
at the time of this study,the survey area had increased to 3577
deg$^2$ from the 2060 deg$^2$ available to \citetalias{Carnall2015},
partly by only using $riz$ band-merged CASU catalogues. This maximises
the searchable area because the $u$ and $g$ sky coverage is less than
for $riz$. Further, the CASU database has higher sky coverage than the
public accessible Edinburgh Wide Field Astronomy Unit (WFAU) archive. The
average $5\sigma$ limits of the ATLAS $riz$ catalogue for point sources
as measured in a $1''$ radius aperture are $r<22.7$, $i<22.0$ and
$z<20.9$ mag. See \cite{Shanks2016} for a comparison of  ATLAS 
parameters with various other surveys.

The main issues with the CASU catalogues are that they are currently CCD
based rather than tile based (as at the WFAU). This leads to increased
incompleteness in the matching at the edges which increases the numbers
of spurious $r$- and $i$-band dropouts.

\subsection{WISE}

The Wide-field Infrared Survey Explorer (WISE, \citealt{Wright2010})
survey covers the mid-IR 3.4 (W1), 4.6 (W2), 12 (W3) and 22 (W4) micron
bands. The advantage of WISE is its 100\% coverage of ATLAS at the
present time and the excellent matching of both the W1 and W2 bands'
depths to VST ATLAS. The approximate $5\sigma$ limits for
AllWISE\footnote{http://wise2.ipac.caltech.edu/docs/release/allwise/}
point sources are $W1=16.90$ and $W2=15.95$ mag in the Vega system. The
W1 and W2 bands have point spread functions (PSFs) of $6.''1$ and
$6.''4$ respectively, compared with $\approx0.''85$ in the VST ATLAS
$riz$ bands. Astrometric
tests\footnote{http://wise2.ipac.caltech.edu/docs/release/allsky/expsup/sec2\_2.html} 
between WISE and  USNO CCD Astrograph Catalog (UCAC3) positions shows
$<0.''5$ rms offset between the two catalogues to the W1 limit. ATLAS
optical photometry was matched to the publicly available ALLWISE
Catalogue using a $3''$ matching radius. For the sky density of WISE
sources at $|b|>30^{\circ}$ we calculate that $\approx4$\% of candidates
identified in WISE will have a blended WISE source within $3''$.
In looking for rare, high-$z$, quasars, WISE blends (and other
artefacts) will therefore have to be eliminated by visual inspection.

\begin{table*}
\begin{center}
\begin{tabular}{ccccc}
\hline
 Quasar            & Redshift & Observation Date     & ESO Project No. &  DIMM Seeing\\
\hline
 VST-ATLAS J025.6821-33.4627 & 6.31 $\pm$  0.03 & 19/11/15  &  096.A-0418(A) & $1.''2$\\
 VST-ATLAS J029.9915-36.5658 & 6.02 $\pm$  0.03 & 22/02/15  &  294.A-5031(B) & $1.''2$\\
 VST-ATLAS J332.8017-32.1036 & 6.32 $\pm$  0.03 & 24/10/15  &  096.A-0418(A) & $1.''5$\\
 VST-ATLAS J158.6938-14.4211 & 6.07 $\pm$  0.03 & 23/01/16  &  096.A-0418(B) & $1.''4$\\
\hline
\end{tabular}
\end{center}
\caption{Details for VLT X-shooter observations. DIMM refers to
Differential Image Motion Monitor atmospheric seeing estimates from ESO.
Redshifts were estimated from the wavelengths of the
Ly$\alpha$ emission line the X-shooter spectra. For J158-14 and J332-32
these differed slightly from the discovery spectrum redshifts reported
in Sections \ref{sect:J158} and \ref{sect:J332}. 
}
\label{table:xshooter}
\end{table*}

\section{Photometric Quasar Selection}
\label{quasarselection}

\citetalias{Carnall2015} used ATLAS $i-z$ colour to select dropout
candidates, then the $z-W2:W1-W2$ colour plane to discriminate between
dwarf stars and high redshift quasars, as shown in Figure
\ref{fig:colselection}. $W1-W2$ colour is the main discriminator against
L dwarfs and $z-W2$ is used to discriminate between T dwarfs and
quasars, providing good separation out to $z=6.3$ before the quasar
colour track crosses the T dwarf locus. Application of this technique to
the first 2060 deg$^2$ of VST ATLAS imaging resulted in the discovery of
two $z>6$ quasars, ATLAS J029.9915-36.5658 at $z=6.02\pm0.03$ and ATLAS
J025.6821-33.4627 at $z=6.31\pm0.03$ and the re-discovery of VIKING/KiDS
J0328-3253 at $z=5.86\pm0.03$.

Following \citetalias{Carnall2015} we first produced a list of high
redshift ($z>5.7$) quasar candidates by applying the following selection
criteria to the CASU band-merged catalogues:

\medskip

\noindent \textbf{(i)} Objects must have $18<z_{AB}<20$ mag, measured
in	a $1''$ aperture (labelled as {\tt aper3} in the CASU and WFAU
catalogues). \cite{Chehade2016T} shows that this aperture provides the
best balance between $S/N$ and accuracy of the aperture corrections for
point sources. The $z_{AB}>18$ cut reduces the number of candidates for 
visual inspection in a range where the star /quasar number ratio 
is expected to increase.

\noindent \textbf{(ii)} Objects must be identified as a point source by
their curve of growth (see \citealt{Gonzalez2008}). Morphology is stored
under the flag {\tt Classification} where $-1$ is for point sources and
$+1$ is for extended sources. We disregard morphological classification
in the $i$-band, since at low $S/N$, the {\tt Classification} is
unreliable. \cite{Chehade2016} tests the CASU morphological
classification and finds it to be $\approx\ 90\%$ complete for the $g$
and $r$ bands. We examine the $z$-band specifically in Section
\ref{pointcompleteness}.

\noindent \textbf{(iii)} Objects must either be undetected in the
$i$-band, or if detected must have $i-z>2.2$. An additional
selection was also carried out for which this limit was relaxed to
$i-z>1.8$, to fill observing gaps between higher priority targets.
Objects must also be undetected in the $u$, $g$ and $r$ bands.

\noindent \textbf{(iv)} Objects must have a corresponding source in the
ALLWISE catalogue within a $3''$ radius with $\frac{S}{N}>3$ in both the
W1 ($\lesssim17.9$ mag) and W2 ($\lesssim17.0$ mag) bands. 

\noindent \textbf{(v)} Objects must have colours in the range
$0.55<W1-W2<1.2$ and $3.2<z_{AB} - W2<4.5$ (see Figure
\ref{fig:colselection}).

\medskip

\noindent This left $\approx130$ candidates for visual inspection. The
candidates were largely obviously misidentified as dropouts due to
defects in the band-merging process, or in areas where the $i-$band
seeing was poor ($>1.''2$), meaning the dropout status could not be
confirmed due to a lower limiting colour. We also limited our selection
regions to the doubly exposed regions of the VST ATLAS stack (see
\citealt{Shanks2015}) for this reason. This reduced the searchable area
per tile from $\sim1$ deg$^2$ to $\sim0.8$ deg$^2$.

As well as our primary selection, secondary targets were identified in
the bluer, $0.3<W1-W2<0.55$ colour region and tertiary targets in the
$0.3<W1-W2<0.55$ and $1.8<i-z<2.2$ region of colour space. The bluer
WISE cut was motivated by the presence of two previously discovered SDSS
quasars of this colour which our original selection criteria of
\citetalias{Carnall2015} miss. Whilst the region lies closer to the L
dwarf locus it provided a test of the incompleteness of our more
conservative $W1-W2$ selection.

In total, we identified two candidates of highest priority within our
primary selection region. These candidates were identified as highest
priority due to their high signal-to-noise ratio (SNR) detections. A
further seven primary candidates in areas of higher background noise
and/or near CCD edges were also identified, along with ten secondary and
five tertiary candidates in the VST ATLAS footprint, as observable from
La Silla in our May 2015 NTT EFOSC2 observing run (see Section \ref
{sect:efosc2} below).


\begin{figure*}
\includegraphics[width=0.7\textwidth,angle=-90]{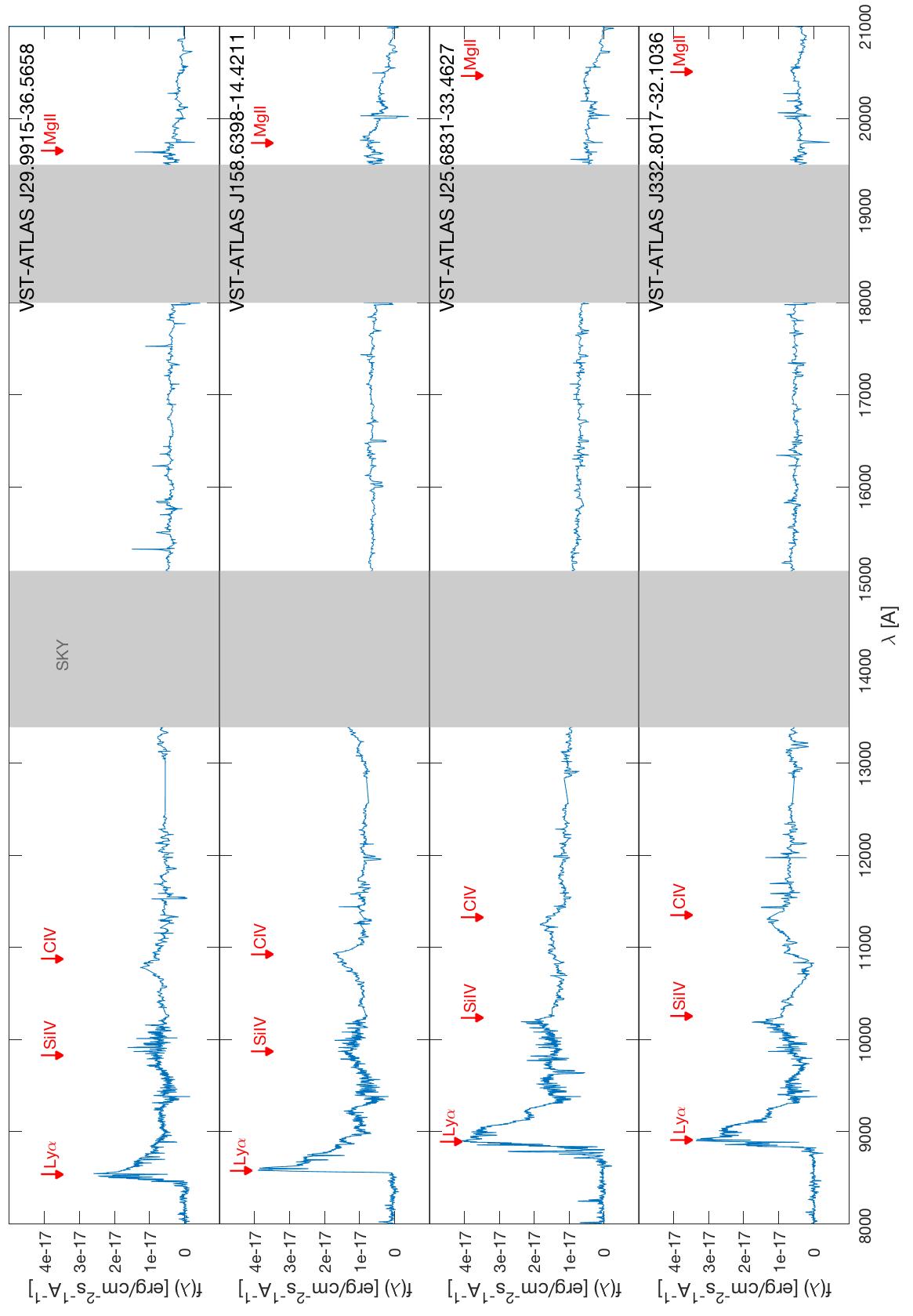}
\caption{The VLT X-shooter spectra for our 4 ATLAS quasars. (See Table
\ref{table:xshooter} for observational details). The spectra
were flux calibrated to the observed $z$-band magnitudes from VST ATLAS
and corrected for telluric absorption as described in Section \ref{sect:xshooter}. 
The positions of the Ly$\alpha$, Si IV, C IV and MgII 
emission lines are marked at the redshifts given in Table
\ref{table:xshooter}. We note that higher ionisation lines such as C IV
are frequently found to be blueshifted with respect to lower ionisation
lines like Ly$\alpha$ and this is seen to be the case for all four
quasars.}
\label{fig:xshooter}
\end{figure*}

\section{Follow-up Spectroscopy}
\label{spectroscopy}

\subsection{Keck LRIS Spectroscopy}
\subsubsection{VST-ATLAS J158.6938-14.4211}
\label{sect:J158}

On 2015 April 19 we observed one of our two highest priority candidates
with the Low Resolution Imaging Spectrometer
\citep[LRIS][]{Oke1995,Rockosi2010} on the Keck I telescope. The
observers were X. Prochaska (P.I.), J. Hennawi and C. Leibler. The
target was VST-ATLAS J158.6938-14.4211, which was observed for 600 s
through a $1''$ wide slit. We used the 400/8500 grating in the red arm
giving $\approx6.9\angstrom$ resolution and a dispersion of
$1.16\angstrom {/}\mathrm{pixel}$ and the 400/3400 grism in the blue arm
giving $\approx7\angstrom$ resolution and a dispersion of $1.09
\angstrom{/}\mathrm{pixel}$.  The LRIS spectrum of J158.6938-14.4211 was
reduced with the LowRedux pipeline\footnote{http://www.ucolick.org/\textasciitilde{}xavier/LowRedux/index.html}.
First, the pipeline processes the calibrations (bias, flat fields and sky
flats) and computes the wavelength solution using arc lamps. Next, these
calibrations are applied to the raw science frames, and a 1D spectrum is
extracted from the 2D frames. The spectra are then flux calibrated using
observations of a spectrophotometric standard star. The LRIS spectrum
showed strong Ly$\alpha$ emission, cut off sharply towards blue
wavelengths due to a strong Gunn-Peterson hydrogen absorption trough,
confirming this as a quasar with an approximate redshift of
$z=6.05\pm0.03$ (but see Section \ref{sect:xshooter} and Table
\ref{table:xshooter}). Also detected in emission are Ly$\beta$, NV(1240
\angstrom), OI(1304 \angstrom) and SiIV(1398 \angstrom). Other details
of this quasar, including $i$, $z$, $W1$ and $W2$ magnitudes are given
in Table \ref{table:table1}. 

\begin{figure*}
\includegraphics[width=0.8\textwidth]{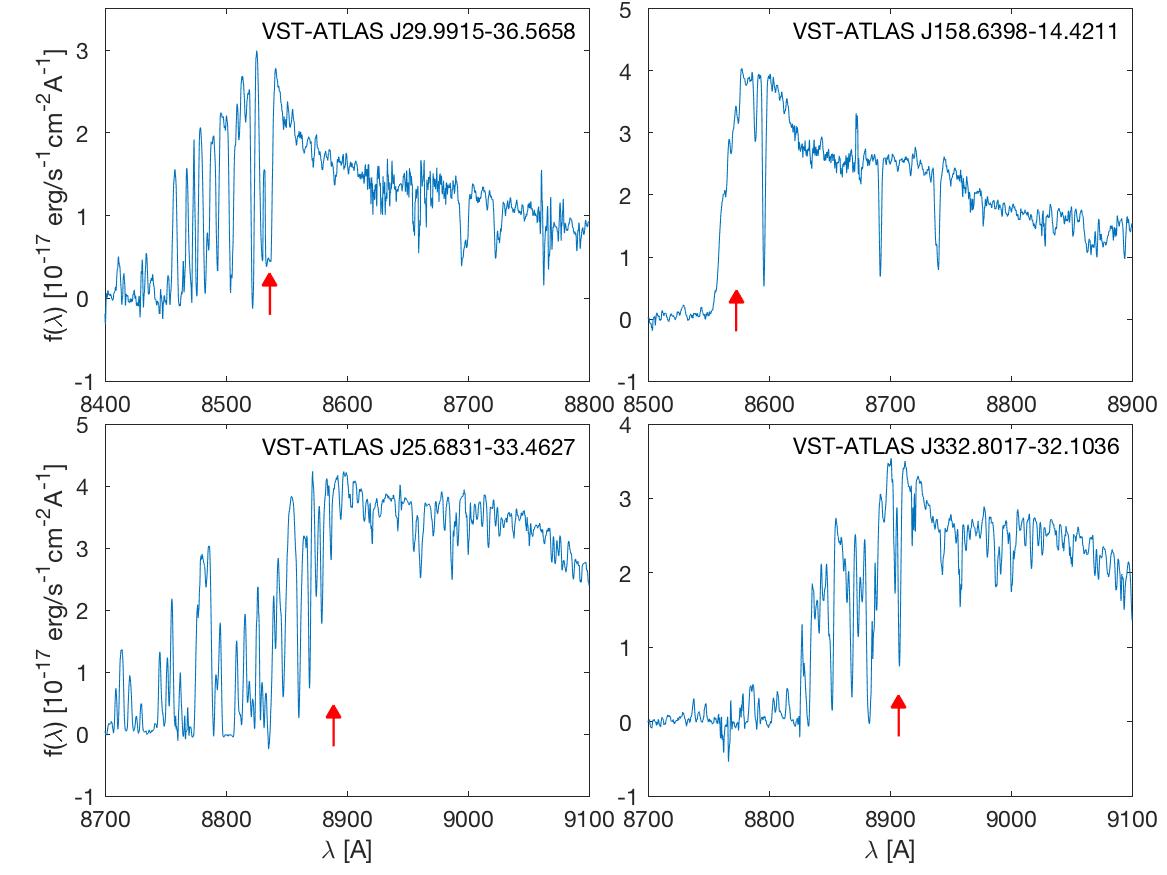}
\caption{The VLT X-shooter spectra for our 4 ATLAS quasars now `zoomed in' to emphasise the 
HI absorption structure around the Ly$\alpha$ emission line. The positions of the Ly$\alpha$ 
emission lines as estimated from the maxima of the line profiles are indicated by the red arrows.
 }
\label{fig:xshooterLyA}
\end{figure*}

\subsection{ESO NTT/EFOSC$2$ Spectroscopy}
\label{sect:efosc2}
\subsubsection{VST-ATLAS J332.8017-32.1036}
\label{sect:J332}
On 2015 May 29 we observed our second highest priority candidate with
the European Southern Observatory's Faint Object Spectrograph and Camera
2 \citep[EFOSC2,][]{Buzzoni1984} on the 3.58m ESO New Technology
Telescope (NTT). The observers were B. Chehade and T. Shanks and the
target was VST-ATLAS J332.8017-32.1036. The target was observed for
$600s$ through Grism No. 2 (100 lines/mm) and a $1''$ wide slit giving a
resolution of 49.6\angstrom\ (FWHM). The pixels were binned $2{\times}2$
in order to reduce readout time and noise. The resulting spatial scale
was $0.25''$/pixel and the dispersion was 13.2\angstrom/pixel. All
stages of the data reduction were performed using standard {\tt \sc
IRAF} routines. The wavelength calibration was performed against
Helium-Argon (HeAr) arc lamp spectra. The spectra cover the wavelength
range of $\approx5100{-}11000\angstrom$. Observations of the
spectrophotometric standard star LTT3864 \citep{Hamuy1992,Hamuy1994}
were used for absolute flux calibration.

As with VST-ATLAS J158.6938-14.4211, strong Ly$\alpha$ emission was seen
redwards of a Gunn-Peterson absorption trough, again confirming that
this object was correctly identifed as a high redshift quasar with
approximate redshift of $z=6.37\pm0.03$ (but see Section
\ref{sect:xshooter} and Table \ref{table:xshooter}). The NV
(1240\angstrom) emission line might also be detected at 9140\angstrom.
There is also a hint of Ly$\beta$ emission at 6520\angstrom. Further
discussion of these spectral features is postponed to Section
\ref{sect:xshooter}, in which we report higher resolution X-shooter
spectroscopy. Magnitudes for this quasar are also reported in Table
\ref{table:table1}.

\subsubsection{PSO J340.2041-18.6621}
\begin{figure*}
 	\includegraphics[width=0.9\textwidth]{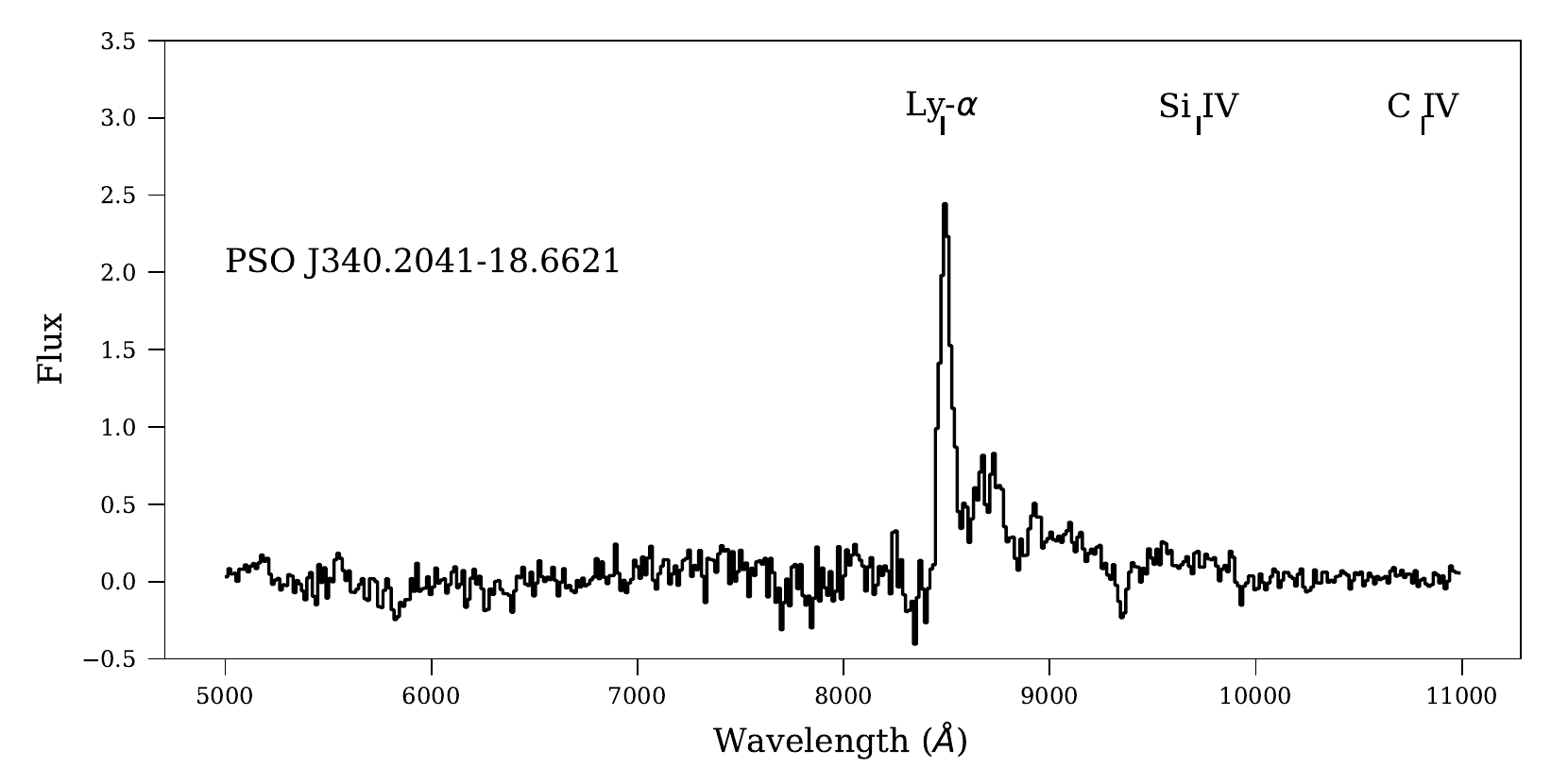}
 	\caption{NTT/EFOSC2 spectrum for the re-discovered
 	quasar PSO J340.2041-18.6621 of \protect\cite{Banados2014}.  The positions 
 	of $Ly\alpha$, $SiIV$ and $CIV$ emission lines at $z=5.98$ are shown.}
 	\label{fig:nttz5pt98}
\end{figure*}

A third high redshift quasar was observed with NTT EFOSC-2 on the night
of 2015 June 01. Using the same VST ATLAS$+$WISE combined selection
technique, we re-discovered the $z=5.98$ quasar, PSO J340.2041-18.6621,
independently found by \cite{Banados2014} using the PanSTARRS survey.
The spectrum confirmed the object as a quasar although we measured  a
slightly lower redshift of $z = 5.98\pm0.03$, based on the Ly$\alpha$
emission line, than that measured by \cite{Banados2014} of $z = 6.03$.
We include this quasar in Table \ref{table:table1} for completeness and 
its NTT spectrum is shown in Fig. \ref{fig:nttz5pt98}.

\subsection{VLT X-shooter Spectroscopy}
\label{sect:xshooter}
To increase the wavelength coverage for the spectroscopically confirmed
quasars from this work and \citetalias{Carnall2015} we observed the four
VST ATLAS quasars with the medium resolution spectrograph X-shooter
\citep{Vernet2011} on the Cassegrain focus of the 8.2m VLT Kueyen (UT2)
on the dates shown in Table \ref{table:xshooter}. The  X-shooter
spectra are shown in Fig. \ref{fig:xshooter}.

The three arms of the X-shooter instrument are the UVB
($3000-5595\angstrom$), the VIS ($5595-10240\angstrom$) and the NIR
($10240-24800\angstrom$). Given the redshift of our targets, we
disregard the UVB arm data in our present analysis. The slit width for
both the VIS and NIR arms was $0.''9$ and slit length was $11''$. Pixel
scales are $\sim0.15\angstrom/\mathrm{pixel}$ and
$\sim0.2\angstrom/\mathrm{pixel}$ for the VIS and NIR arms respectively.
The resolution of the observations were $R\approx7400$ in the VIS arm
and $R\approx5400$ in the NIR arm. We observed each of the 4 targets for
1hr each, $\approx45$ mins of which was on-sky exposure. The X-shooter 
spectra are shown in Fig. \ref{fig:xshooter}.

To reduce the X-shooter data, we built a tailor made workflow that
incorporates both standard ESO recipes but also IRAF tasks to improve
data quality. The ESO pipeline was used to perform the reduction steps
up to the 2D rectified and merged spectrum. This includes bias
subtraction, flat-fielding, modelling of the spectral orders, resampling
and rectifying the spectra. We ensured that the default pipeline
parameters were producing optimal results and adjusted them in a few
cases, namely cosmic ray removal and sky background subtraction. To
extract the QSO 1D spectra from the 2D spectra we used the IRAF task
{\it apall}. Finally, we applied a telluric correction to the NIR arm
and merged the VIS and NIR arms to produce the spectra displayed in Fig.
\ref{fig:xshooter}. Regions with transmission of less than $\approx20$\%
were masked from the spectra. 

We show the  positions of the Ly$\alpha$,
SiIV, CIV and MgII emission lines at the redshifts in Table
\ref{table:xshooter}; these are based on the X-shooter redshifts
estimated from the position of Ly$\alpha$ indicated in Fig.
\ref{fig:xshooterLyA}. \footnote{We prefer these Ly$\alpha$ redshifts rather than
e.g. the MgII emission line redshifts given in Table \ref{table:mbh} below on
the grounds of the higher S/N of the Ly$\alpha$ line.}  Note the slight
differences in these redshifts from those measured in the discovery
spectra (see Table \ref{table:table1} and Table 1 of
\citetalias{Carnall2015}). The higher ionisation lines such as C~IV are
frequently seen to be blueshifted with respect to the lower ionisation
lines (e.g \citealt{Coatman2016}) and this appears to be the case for
these four quasars. However, it is not ruled out that there could be
some uncertainty in our redshifts due to variation in the Ly$\alpha$
emission line profile. These issues are further discussed in Section
\ref{sect:mbh} where redshifts are estimated from the CIV and MgII
emission lines. The CIII] emission lines lie inside the range affected
by atmospheric absorption in  the X-shooter NIR data and so are
unavailable to further test our redshifts. We note that quasar
J332.8017-32.1036 may also show broad absorption features below the Si
IV and C IV emission lines i.e. it may be a broad absorption line quasar
(BALQ). We used these X-shooter spectra to measure $M_{1450\angstrom}$
absolute magnitudes, as reported in Table \ref{table:absmag}. We also
make a preliminary calculation of black hole masses in Section
\ref{sect:mbh} but defer the detailed analysis and the fitting of
Ly$\alpha$ emission and associated absorption line profiles (see Fig.
\ref{fig:xshooterLyA}) to future work (Diener et al., in prep.).

\subsection{Other candidates and effect of relaxing the W1-W2 limit}

Our original selection criteria as listed in Section
\ref{quasarselection} have produced the discoveries of four $z>6$
quasars from the top of the ATLAS$+$WISE priority list. However, none of
the other candidates which were also observed as part of our four night
NTT run between 2015 May 29 and June 01, including our sample selected
to be bluer in $W1-W2$ and $i-z$, displayed spectral features consistent
with high redshift quasars. These candidates had  been marked as lower
priority due to their lower signal to noise detections meaning a larger
probability of these objects having scattered into our selection region
from the stellar locus (see Fig. \ref{fig:colselection}). We
subsequently re-measured the $i$-band photometry for the $i$-band
dropouts based in the $z$-band detections using the {\tt \sc IMCORE}
routines supplied by CASU. The non-quasar targets were generally positively 
detected at below the $5\sigma$ limit of the ATLAS catalogue i.e. they were not $i$-dropouts.


\section{Consistency with previous quasar luminosity function estimates}
\label{qlf}
In this Section we will use our discovery of four (plus two
re-discovered) $z>6$ quasars from \citetalias{Carnall2015} and this work
to check consistency with previous studies of the quasar luminosity
function at bright absolute  magnitudes. We estimate the completeness of
our catalogues with respect to their depth and our ability to identify
point-sources. Next, we test the completeness of our colour selections
from Section \ref{quasarselection}. Finally, we check  our completeness
estimates against the Pan-STARRS survey of \cite{Banados2016} before
comparing our quasar space density to  that expected from previous
estimates of the quasar luminosity function.

\subsection{Catalogue depths}
To determine the completeness of VST ATLAS catalogues as a function of
magnitude we need to compare the number of sources to the expected
number of sources. We take advantage of the overlap between VST ATLAS
and the PanSTARRS Medium Deep Survey (MDS, \citealt{Chambers2016}). The MD02
field of the MDS survey is covered by the VST ATLAS survey. The MD02
centre is RA 03:32:24, Dec -28:08:00 with a radius of $\approx1.5$deg. The
approximate $5\sigma$ limits for this field in $riz$ are 25.4, 25.8 and
25.3 mag  (c.f. 22.7, 22.0 and 20.9 mag for VST ATLAS).

We wish to compare VST ATLAS exposures that are representative of the
survey. To do this, we select a single VSTATLAS tile in each band with
seeing similar to median seeing for the survey ($0.''91$, $0.''69$, $0.''85$ for
the $riz$ bands respectively).

Firstly, we assume that the MD02 field is complete and uniform to the
depth that we test to (24th, 23rd and 22nd magnitude [AB] for the $riz$
bands respectively). Given the dither pattern of the VST ATLAS survey
different areas in the stack will either have two, one or no exposures.
For each detection brighter than the faint limit in the MDS catalogue we
compare to the VST ATLAS images, masking the MDS data according to the
depth of VST coverage.

We compared {\tt aper3} magnitudes from VST ATLAS to the Panstarrs PSF
magnitudes and found offsets of $+0.06\pm0.03$, $-0.19\pm0.02$ and $+0.08\pm0.03$ mag in
the $riz$-bands. After correcting for these differences we can check
VST ATLAS catalogue completeness.

\begin{figure}
	\includegraphics[width=0.45\textwidth]{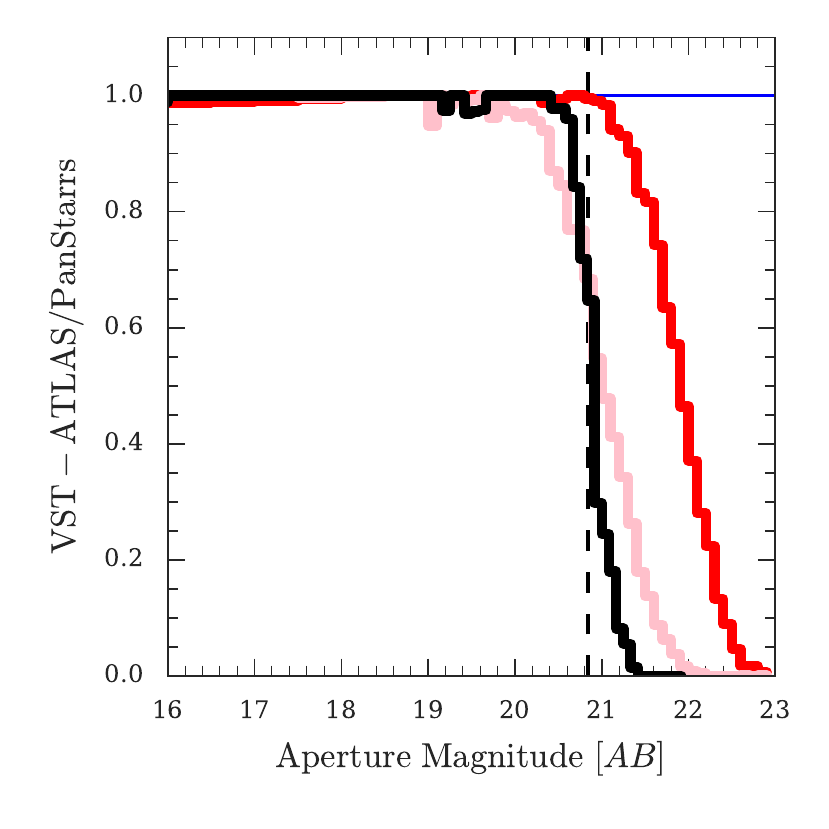}
	\caption{Completeness of the $r$, $i$ and $z$-bands (red, pink and
	black respectively)  as a function of magnitude. Fractions are
	derived by comparing catalogue detections in VST ATLAS with
	detections in the Panstarrs MD02 field. The vertical dashed line
	corresponds to the $5\sigma$ detection limit for the VST ATLAS
	$z$-band.  The catalogue is 99.6\% complete down to $z_{AB}=20.0$
	mag.}
	\label{fig:depth}
\end{figure}

In Fig. \ref{fig:depth} we show the completeness of VST ATLAS
catalogues compared to the Panstarrs MDS as a function of magnitude. We
find that  the catalogue is $99.6$\% complete to $z_{AB}=20$, assuming
that PanSTARRS is 100\% complete..

\subsection{Point-source completeness}
\label{pointcompleteness}
To remove contamination of our colour selection from local galaxies we
select only point-sources as determined by their single band ($z$)
morphological classification in the CASU catalogues. 
To test the completeness of the CASU morphological classification we modelled
the difference between Petrosian and {\tt aper1} ($0.''5$ radius)
aperture magnitudes in the $z$-band for galaxies. We identified galaxies
as all objects being $0.4$mag brighter in Petrosian than in the
{\tt aper1}  magnitude. We fit the distribution with a 
Gaussian model, allowing the centre, width and height to vary. We did
this for all targets within four magnitude ranges; $18<z<18.5$,
$18.5<z<19$, $19<z<19.5$ and $19.5<z<20$ mag. We excluded targets
identified as noise (by their CASU {\tt Classification}), targets in the
singly exposed region and targets in regions with poorly fit sky
background and de-blended photometry.

\begin{figure}
 	\includegraphics[width=\linewidth]{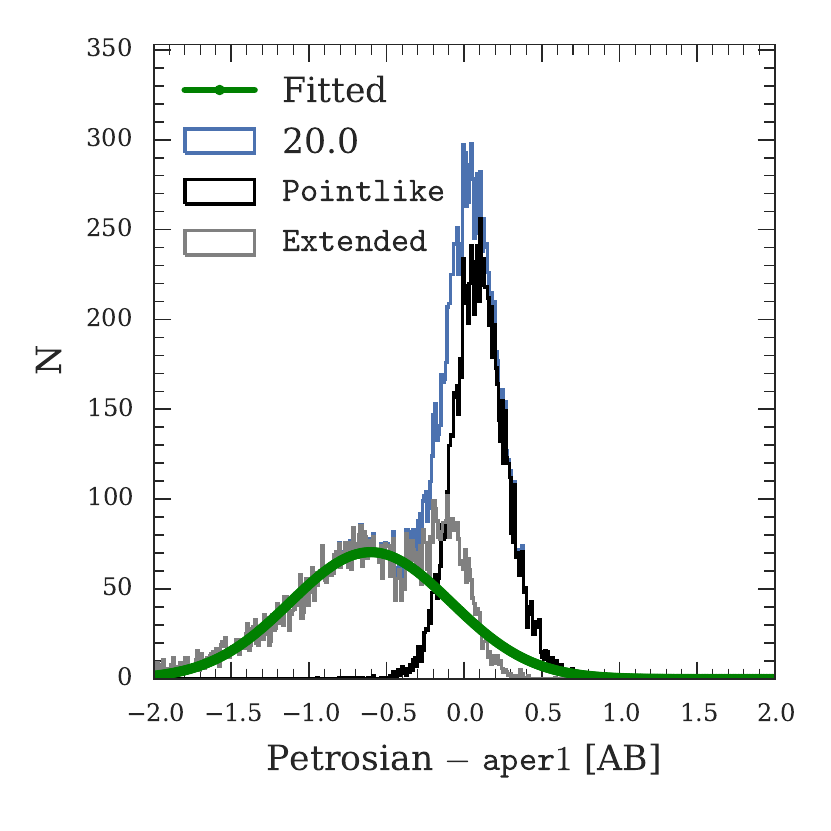}
	\caption{We show the distribution of Petrosian-{\tt aper1} ($0.''5$
	radius aperture) magnitudes in the $z$-band. The data consists of a
	single concatenation of seventeen tiles. The images were taken on
	night 2014-06-17 with median $0.''86$ seeing (c.f. $0.''84$ median
	for the survey). We plot the total histogram of objects (blue)
	between $19.5<z_{AB}<20.0$ mag i.e. the faintest 0.5 mag of our
	target candidates. The fitted galaxy distribution is shown as a
	solid green line. The  black and grey histograms show the point-like
	and extended sources (as identified by CASU) respectively.}
	\label{chpt2:fig:zmorph}
\end{figure}

Fig. \ref{chpt2:fig:zmorph} shows the performance of the morphological
separation of stars and galaxies in our faintest $19.5<z_{AB}<20.0$ mag
range. Subtracting the modelled distribution of galaxies in the
Petrosian-{\tt aper1} plane to the distribution of all objects we
may estimate the number of stars not identified as point-sources by
CASU. We find the predicted extra number of stars in the above four 
bins (brightest to faintest) of 416, 837, 1277 and 704, implying
94\%, 90\%, 87\% and 93\% point-source completeness respectively.
Increasing the aperture size to {\tt aper2}  ($0.''7$ radius) yields similar results for
$z<19$ but the separation in the Petrosian-{\tt aper2} plane is
less well defined so poorly fits the distribution of galaxies. Including
these `missed' stars to those identified by CASU we find an increase in
the number of stars of 10\%. Based on this analysis we estimate that
our point-source completeness is 91\%. This number agrees well with
the estimate of \cite{Chehade2016} where stars were colour selected
using multiple VST ATLAS and WISE bands.

\subsection{Colour completeness}
There are two methods used by other $z\sim6$ quasar searches to
calculate the incompleteness of their samples. The first is to clone low
redshift quasar spectra and shift them to higher redshifts, \citep[see
][]{Willott2005,Findlay2012}. The second method, which we use in this work, is to
create model quasar spectra based on empirical models of quasar SEDs
\citep{Fan1999,McGreer2013}. Both methods assume that the low redshift
UV properties of quasars, those  redward of Ly${\alpha}$, do not evolve
with redshift \citep{Fan2004,Jiang2006}.  

To generate our model quasar spectra we use the publicly available {\sc
\tt SIMQSO}\footnote{https://github.com/imcgreer/simqso} code developed
by \cite{McGreer2013}. In this code, each quasar is assigned a power law
continuum with a break at $1100$\AA. The slopes are drawn from normal
distributions based on the results of \cite{Telfer2002}. Emission lines
with Gaussian shape are added to this continuum and the emission line
properties (wavelength, equivalent width and FWHM) are also drawn from
normal distributions. The emission lines distributions are generated
from composite BOSS spectra which have been stacked in different
luminosity bins. The emission blueward of Ly$\alpha$ is based on the
work by \cite{Worseck2011}. The model relies on the observed number
densities of high column density systems \citep{Songaila2010}. A large
number of sightlines are generated and the mean free paths are estimated
by matching to the observations of \cite{Songaila2010}. The column
density distribution function is used to estimate the effective optical
depth ($\tau_{\mathrm{eff}}$) which is checked for consistency with
observations \citep{Songaila2004,Fan2006}. Further details of the model
may be found in \cite{McGreer2013}. Note that we  use  the luminosity
function parameters of \cite{Willott2010a}. To accommodate WISE colour
selection at higher redshifts we follow \cite{Yang2016}, adding three new spectral 
breaks at 5700, 10850, and 22300\angstrom\, and including their assumed spectral slopes and dispersions.
Emission line parameters are derived from the composite quasar spectrum of \cite{Glikman2006}.
 
\begin{figure}
	\includegraphics[width=0.45\textwidth]{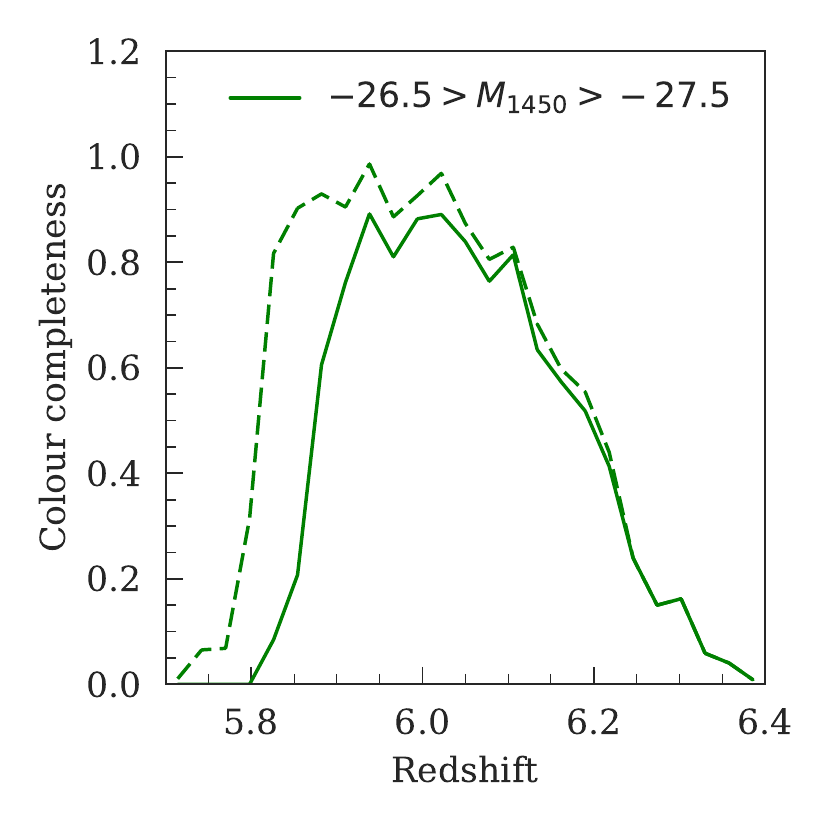}
	\caption{Colour completeness as a function of redshift. The solid
	line shows the completeness when we apply the most conservative
	selection from Section \ref{quasarselection} and the dashed line shows the
	completeness using the relaxed selection criteria.}
	\label{fig:colourcompleteness}
\end{figure}

We use {\sc \tt SIMQSO} to generate $140,000$ model quasars evenly
distributed between $5.7<z<6.4$ and $-26.5\ge\mathrm{M}_{1450\angstrom}\ge-27.5$.
In Fig. \ref{fig:colourcompleteness} we show the resulting colour
completeness using the photometric selections from Section
\ref{quasarselection}. The completeness of the conservative selection
rises sharply between redshifts $5.7<z<5.9$ due to the $i-z$ colour
selection. When the colour selection is relaxed (shown by the dashed
line) we see that the colour completeness is increased at lower
redshifts. At higher redshifts ($z\gtrsim6.05$) the completeness falls
due to the faint limit of $z_{AB}=20$ mag.


It should be noted that the 6 quasars in Table \ref{table:absmag} show a
redshift  distribution apparently skewed to higher redshifts than
implied by Fig. \ref{fig:colourcompleteness}. Indeed, the {\sc\tt
SIMQSO} quasar count model  predicts that $\sim2/3$ of detected quasars should
lie in the $5.7<z<6.0$ range whereas only $\sim1/3$ of those in Table
\ref{table:absmag} do. Indeed, the quasars selected by \cite{Banados2016}
are more consistent with this prediction. However, lacking any other
explanation, we note that this larger than expected $z>6$ fraction for
ATLAS at least remains within the bounds of statistical error.

\subsection{Completeness estimates  checked via PanSTARRs}

\cite{Banados2016} have searched for $5.7\le z\le6.7$ quasars  using the
PanSTARRS1 (PS1) $3\pi$ survey. This survey covers the ATLAS sky
footprint except for the areas below $Dec<-30$ deg in the SGC and NGC
areas. \cite{Chambers2016} indicates that PS1 reaches $\approx1.5$ mag fainter than
ATLAS in $z$, with a 5$\sigma$ stellar limit of $z_{AB}=22.3$ mag as
opposed to $z_{AB}=20.87$ mag for ATLAS. \cite{Banados2016} quote a median
limit of $z=22.0$ mag for their PS1 survey. 5 extra quasars were detected in
the ATLAS overlap area with $z_{AB}\le20.0$ mag; 4 of these had $z<6$. One
quasar was missed by ATLAS because it was classed as a galaxy (PSO
J056.7168-16.4769), making an incompleteness of 1/7 or $\approx14$\% of the 7
quasars that were detected by either survey in the ATLAS-PS1 overlap
area. This is similar to  our 10\% star-galaxy incompleteness estimate.
One quasar was missed on account of its $i-z$ colour being too blue (PSO
J021.4213-25.8822) and two because either their $W1-W2$ colour was too
blue (PSO J183.2991-12.7676) or there was no WISE counterpart within
$3''$ ( CFHQS J1509-1749). This makes the colour incompleteness 3/7 or
$\approx43$\% compared to our incompleteness estimates of between 10-60\%
depending on redshift (see Fig. \ref {fig:colourcompleteness}). We
conclude that the extra quasars detected by \cite{Banados2016} are not
inconsistent with our star-galaxy separation and colour incompleteness
estimates, although clearly there may still be further residual
incompleteness associated with the PS1 survey.


\begin{table*}
\begin{center}
\begin{tabular}{ccccc}
\hline
 Quasar            & Redshift  & $z$ (AB mag)     & $M_{1450\angstrom}$ &  Reference\\
\hline
 ATLAS J025.6821-33.4627 & 6.31 $\pm$  0.03 & 19.63 $\pm$ 0.06 &  -27.50 $\pm$ 0.06& \cite{Carnall2015}\\
 ATLAS J029.9915-36.5658 & 6.02 $\pm$  0.03 & 19.54 $\pm$ 0.08 &  -26.97 $\pm$ 0.08& \cite{Carnall2015}\\
 VIKINGKiDS J0328-3253   & 5.86 $\pm$  0.03 & 19.75 $\pm$ 0.12 &  -26.60 $\pm$ 0.04& \cite{Venemans2015b}\\
 ATLAS J332.8017-32.1036 & 6.32 $\pm$  0.03 & 19.75 $\pm$ 0.06 &  -26.79 $\pm$ 0.06& This paper \\
 ATLAS J158.6938-14.4211 & 6.07 $\pm$  0.03 & 19.44 $\pm$ 0.08 &  -27.23 $\pm$ 0.08& This paper \\
 PSO J340.2041-18.6621   & 5.98 $\pm$  0.03 & 19.67 $\pm$ 0.10 &  -26.42 $\pm$ 0.10& \cite{Banados2014}\\
\hline
\end{tabular}
\end{center}
\caption{Absolute magnitudes for the  four quasars discovered and the two quasars rediscovered 
in VST ATLAS+WISE. The ATLAS quasar absolute magnitudes are estimated via the X-Shooter spectra in Fig. \ref{fig:xshooter}
and the other two from the above sources.}
\label{table:absmag}
\end{table*}

\subsection{Consistency with previous luminosity function  studies}
\label{sect:qlfcomp}



Following \citetalias{Carnall2015} we calculate $M_{1450\angstrom}$
luminosities by scaling their  SDSS composite  template spectrum from
\cite{vdb2001} and \cite{Songaila2004} to the X-shooter spectra of each
of our four ATLAS quasars in Fig. \ref{fig:xshooter} using their
observed Ly$\alpha$ redshift and $z$-band magnitude. The results are listed in
Table \ref{table:absmag}, where we have also included the absolute
magnitudes for our two re-discovered quasars quoted by
\cite{Venemans2015b} and \cite{Banados2014}. We see that the 6 quasars
occur in approximately one absolute magnitude bin centred on
$M_{1450\angstrom}\approx-27\pm0.5$.

We have estimated the quasar space density at $M_{1450\angstrom}\approx-27$
based on our observations. We assume a search area of 3119 deg$^2$,
representing 80\% of the 3577deg$^2$ area searched i.e. assuming 20\%
area lost at CCD edges. From Fig. \ref{fig:colourcompleteness} and
assuming the relaxed selection criteria of Section \ref{quasarselection}
we take  the redshift range to be $5.7<z<6.4$. We take the
redshift-dependent colour completeness to be that given in Fig.
\ref{fig:colourcompleteness}, the magnitude completeness at $z_{AB}<20$
to be 100\% and the star-galaxy separation completeness to be 90\%,
implying we need to multiply our 6 detected quasars by a factor of 1.9
to account for these incompletenesses. Thus our corrected quasar count
is now $11.3\pm4.6$. We note that the total quasar count in the overlap
area from \cite{Banados2016} and ATLAS is 7 where ATLAS discovered 2,
giving an ATLAS incompleteness factor of 3.5, compared to our modelled
factor of 1.9. Applying this correction to the 6 ATLAS  quasars   gives
a count of $21\pm8.6$ over 3119 deg$^2$. Although 1 out of 2  quasars
available in Table 2 was missed by \cite{Banados2016}, we neglect this
possible evidence for residual incompleteness in their sample on the
grounds that the statistical error on the corrected count would be
dominant. We conclude that the corrected number of quasars in our
sample ($11.3\pm4.6$) is in reasonable statistical agreement with those of
\cite{Banados2016} in a similar area ($21\pm8.6$). As noted already, most of this
difference is due to ATLAS missing $5.7<z<6$ quasars.

\begin{figure}
\includegraphics[width=\linewidth]{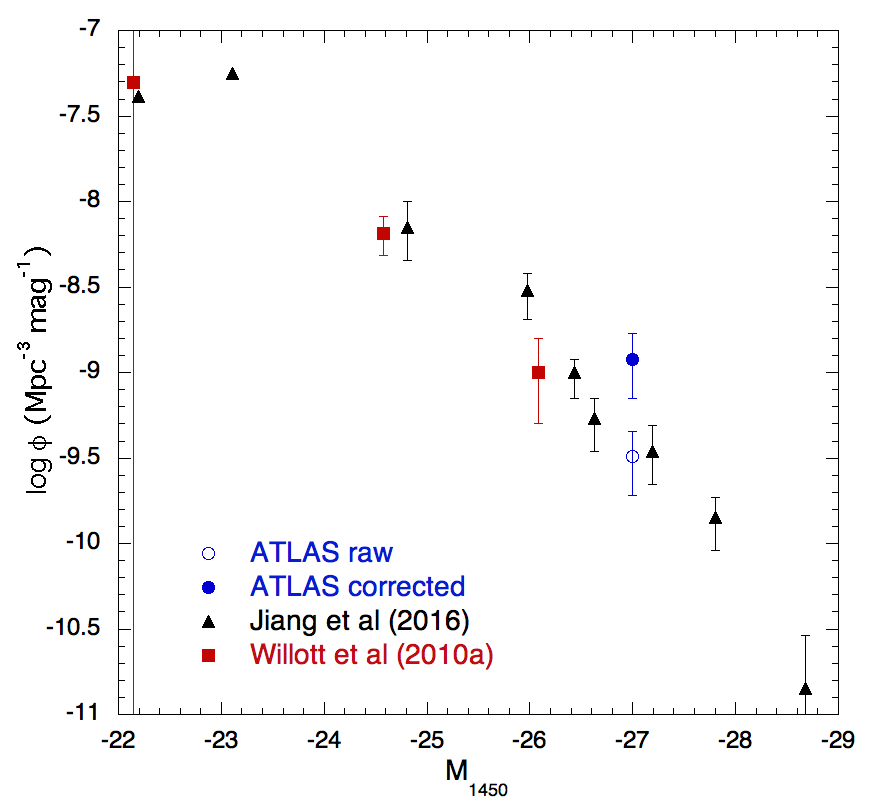}
\caption{Approximate ATLAS quasar space densities at
$M_{1450\angstrom}=-27$ compared to the  SDSS luminosity function   of 
\protect \cite{Jiang2016} and the fainter result of \protect
\cite{Willott2010a}corrected to our assumed cosmology. `ATLAS raw'  refers to six $5.7<z<6.4$ quasars detected
in our 3119 deg$^2$ search and `ATLAS corrected'  refers to a corrected
total of 11.3 quasars in the same volume.}
\label{fig:lf}
\end{figure}

With a 3119 deg$^2$ survey area, our assumed, $\Omega_m=0.3$,
$\Omega_\Lambda=0.7$, $h=0.7$, cosmology indicates  a volume of 17.4
Gpc$^3$ between $5.7<z<6.4$ and assuming a raw count of $6\pm2.4$ and an
incompleteness  corrected count of $11.3\pm4.6$ gives the quasar space
density  estimates shown in Fig. \ref{fig:lf}. Comparing these with the
SDSS luminosity function  of \cite{Jiang2016} and that of
\cite{Willott2010a} at fainter magnitudes, we see there is good
agreement between our observed number of quasars and the SDSS data which
are bracketted by our raw and corrected space density estimates. Although
our PS1 corrected space density would be $\approx2\times$ higher than the corrected 
value shown, its $\approx 40$\% error means there is still no significant 
disagreement with previous luminosity functions.

As an alternative view of how our quasar counts compare to previous
work, we  use the {\sc\tt SIMQSO} code of  \cite{McGreer2013} with the
luminosity function of \cite{Willott2010a} to predict the quasar count in
the  $15<z_{AB}<20$ mag and $5.7<z<6.4$ ranges. The predicted result is
$8.1\pm2.8$ quasars in our 3119 deg$^2$ area. This compares to our raw,
observed count of $6\pm2.4$ and our corrected count of $11.3\pm4.6$
quasars. We conclude that both these estimates are in good statistical
agreement with the predicted value.

\begin{table*}
\begin{center}
\begin{tabular}{cccccccccccccc}
\hline
           &        &               &       &            &CIV(1550\angstrom)&                  	  &                   &&           &       &MgII(2800\angstrom) &                           &\\\cmidrule{3-8} \cmidrule{10-14}
Quasar     &   z    &$\lambda$      &  z    & FWHM       &    v       &$\lambda L_{1350\angstrom}$& $M_{BH}$          &&  $\lambda$&  z    & FWHM               &$\lambda L_{3000\angstrom}$& $M_{BH}$\\
           & (LyA)  &   (\AA)       & (CIV) &(kms$^{-1}$)&(kms$^{-1}$)&$(10^{44}$ergs$^{-1}$)     & ($10^9M_\odot$)   &&  (\AA)    & (MgII)& (kms$^{-1}$)       &($10^{44}$ergs$^{-1}$)    &($10^9M_\odot$)\\
J029-36    & 6.024  & 10804         & 5.970 & 6425       & -2306      &  198                      & 1.4               &&  19677    & 6.027 & 6510               &  440                      & 6.4\\
J158-14    & 6.069  & 10864         & 6.009 & 7895       & -2546      &  318                      & 2.4               &&  19806    & 6.073 & 2434               &  702                      & 1.1\\ 
J025-33    & 6.305  & 11153         & 6.196 &12255       & -4476      &  224                      & 2.2               &&  20422    & 6.294 & 4109               &  448                      & 2.6\\
J332-32    & 6.325  & 11214         & 6.235 &12913       & -3686      &  154                      & 2.7               &&  20516    & 6.327 & 3078               &  300                      & 1.2\\
\hline
\end{tabular}
\end{center}
\caption{Black hole masses estimated from the CIV and MgII broad
emission line widths measured from the X-shooter NIR spectra. CIV
and MgII wavelengths  and FWHM widths are listed. $\lambda
L_{1350\angstrom}$ and $\lambda L_{3000\angstrom}$ monochromatic
continuum luminosities are based on $z_{AB}$ and $K_{AB}$ magnitudes.
The $K_{AB}$ magnitude for J025-33 is estimated from its $z_{AB}$ and
the mean ${(z-K)}_{AB}=0.37$ colour of the other 3 quasars. CIV
estimates of $M_{BH}$ are from eqs. 4, 6 of \protect\cite{Coatman2016}
while MgII estimates follow eq. 1 of \protect\cite{Vestergaard2009}.
Comparing the MgII and CIV values gives an  approximate  error  for the
average $M_{BH}$ of $\pm37$\%.}
\label{table:mbh}
\end{table*}
  
\section{Black Hole Masses}
\label{sect:mbh}

Finally, we have made a preliminary estimate of the black hole masses
powering these quasars. We therefore fitted Gaussian line profiles to the CIV
and MgII lines in the NIR spectra from	X-shooter using the IRAF {\it
splot} routine. Table \ref{table:mbh} shows the measured FWHM and central wavelengths
of the emission lines. As can be seen from Fig. \ref{fig:xshooter}, the
S/N for the CIV line is generally higher than for MgII. The CIV lines can
be seen to be blueshifted with respect to the redshifts based on
the Ly$\alpha$ line as measured from the X-shooter spectra (see Fig. \ref{fig:xshooterLyA}).

We estimated the monochromatic continuum luminosity, $\lambda
L_{1350\angstrom}$, for the CIV estimate from the quasar's  $z_{AB}$
band magnitudes and $\lambda L_{3000\angstrom}$ for the MgII estimate
from their $K_{AB}$ magnitudes. We chose this route rather than using
the X-shooter spectral fluxes from Fig. \ref{fig:xshooter} partly on the
grounds that the overall reliability of the spectrophotometry is
uncertain and partly because the continuum at rest 3000\angstrom~ lies
at longer wavelengths than available from Fig. \ref{fig:xshooter}. If we
assume that the spectrophotometry at 1350\angstrom\, is accurate then  our
$\lambda L_{1350\angstrom}$ estimates in Table \ref{table:mbh} may be too
large by a factor of $\approx2$ due to Ly$\alpha$ emission line contamination,
implying that our resulting CIV $M_{BH}$ estimates may be too large by
$\approx40-50$\%. There may be a similar systematic issue for the MgII $M_{BH}$ estimates.

We then used eq. 4 of \cite{Coatman2016} to correct the
FWHM of the CIV line for effects correlated with the CIV blueshift
relative to lower ionisation lines. 

$$FWHM(CIV,Corr.) = {{FWHM(CIV,Meas.)}\over{(0.41\times CIV_{Blueshift}/1000kms^{-1}+0.62)}}$$

\noindent We then used their eq. 6 to estimate the values of $M_{BH}$ given in Table \ref{table:mbh}.

$$M_{BH} = 10^{6.71}\times {{FWHM(CIV,Corr.)}\over{1000kms^{-1}}}\times ({\lambda L_{1350\angstrom}\over{{10^{44} erg s^{-1}}}})^{0.53}$$

For the MgII based mass estimates we used eq 1 of \cite{Vestergaard2009} in the form:

$$M_{BH} = 10^{6.86} (FWHM(Mg II))/1000)^2 (\lambda L_{3000\angstrom}/10^{44})^{0.5}.$$

We note that the errors on the MgII line widths  will be substantial,
especially for the lower redshift quasars whose lines are close to
strong sky absorption features. The CIV FWHM are better measured but may
have bigger systematic errors due to the large FWHM correction. With
these caveats, the CIV and MgII masses appear to be in reasonable
agreement, lying in the range $1-3\times10^9M_\odot$ for CIV and
$1-6\times10^9M_\odot$ for  MgII. Averaging both values gives 
$M_{BH}=3.9,1.8,2.4,2.0\times10^9M_\odot$ respectively for J029-36,
J158-14, J025-33 and J332-32 and averaging their standard errors gives 
a rough error estimate of $\pm37$\% on these mean values.

These compare to the $(1.24\pm0.19)\times10^{10}M_\odot$ result for the
$z=6.3$ quasar of \cite{Wu2015}, the brightest quasar ao far found at
$z>6$. This quasar has a monochromatic continuum luminosity of $\lambda
L_{3000\angstrom}=(3.15\pm0.47)\times10^{47}$ erg s$^{−1}$ some
$\approx5\times$ brighter than J158-14 which is our brightest quasar by
this measure. J158-14 is $\approx3\times$ less massive. Our four $z>6$
quasars are generally  similar in luminosity and mass to the $z=7.05$
quasar ULAS J1120+0641 \citep{Mortlock2011} and to the $z=6.89$ quasar
J0100+2802 \citep{DeRosa2014}, both with
$M_{BH}\approx2\times10^9M_\odot$. The recently discovered $z=7.54$
quasar \citep{Banados2017} has $M_{BH}\approx7.8\times10^8M_\odot$,
significantly less massive than our four ATLAS quasars, likely due
to lower continuum luminosity, it  being $\approx1$ mag fainter in
apparent $K$ mag. and $\approx0.5$ mag. fainter given its
luminosity distance and assuming zero K-correction.

These ATLAS quasars are therefore close to having some of the most massive
black holes so far discovered and they lie at a redshift where they may
cause significant difficulties due to the fast gravitational growth rates
needed to explain their existence at such early times. A more detailed
$M_{BH}$ analysis, with improved error estimates and, e.g. where  account is
taken of the FeII emission surrounding MgII in measuring its line width,
is postponed to a forthcoming paper.

\section{Conclusions}
\label{conclusions}
Using similar combinations of VST ATLAS and WISE colours as
\citetalias{Carnall2015} we have found a further two $z>6$ quasars and
rediscovered a third at a redshift $z\sim6$. Adding these to the
previous three found by \citetalias{Carnall2015}  makes 6 quasars
discovered in the searched ATLAS area of 3119 deg$^2$, corresponding to
a volume of $\sim17.4$ Gpc$^3$ between $5.7<z<6.4$. We confirmed the two
new quasars using low dispersion spectroscopy at the ESO NTT
and Keck telescopes and have presented X-shooter spectra for these and
the 2 other quasars discovered in ATLAS. The initial criteria listed in
Section \ref{quasarselection} gave a 100\% success rate for these 6
quasars which had the highest priority at the spectroscopic stage,
indicating high selection efficiency. This possibly led to lower
completeness rates and we have estimated the magnitude, star-galaxy
separation and colour completenesses. When the joint effect of these
incompletenesses are taken into account the corrected  number of quasars
rises to $11.3\pm4.6$. This corrected number is reasonably consistent with the
discovery of a further 5 quasars detected in  the overlap between the
ATLAS and PS1 surveys by \cite{Banados2016}. We have compared our
observed quasar space density  at $M_{1450\angstrom}\approx-27.0\pm0.5$ mag
to the SDSS results of \cite{Jiang2016}. Our raw
and completeness corrected space densities  bracket those of
these authors and so can be considered in good agreement with the SDSS
results. Finally, from preliminary virial analyses of the MgII and CIV broad 
emission lines we find black hole masses in the range, 
$M_{BH}\approx1-6\times10^9M_\odot$ for the four ATLAS quasars.

\section{Acknowledgments}
\label{sec:ack}
Based on VST and VLT data products from observations made with ESO Telescopes at the
La Silla Paranal Observatory under program ID's
177.A-3011(A,B,C,D,E.F,G,H,I,J), 294.A-5031(B), 095.A-0506(A) and
096.A-0418(A,B). Observations were also made at the Keck Telescope
(Project U071LA, PI X Prochaska). This publication also makes use of
data products from the Wide-field Infrared Survey Explorer, which is a
joint project of the University of California, Los Angeles, and the Jet
Propulsion Laboratory/California Institute of Technology, and NEOWISE,
which is a project of the Jet Propulsion Laboratory/California Institute
of Technology. WISE and NEOWISE are funded by the National Aeronautics
and Space Administration. 

We also thank the PanSTARRS1 collaboration for
access to the Medium Deep Survey data in advance of publication. The
Pan-STARRS1 Surveys (PS1) have been made possible through contributions
of the Institute for Astronomy, the University of Hawaii, the Pan-STARRS
Project Office, the Max-Planck Society and its participating institutes,
the Max Planck Institute for Astronomy, Heidelberg and the Max Planck
Institute for Extraterrestrial Physics, Garching, The Johns Hopkins
University, Durham University, the University of Edinburgh, Queen's
University Belfast, the Harvard-Smithsonian Center for Astrophysics, the
Las Cumbres Observatory Global Telescope Network Incorporated, the
National Central University of Taiwan, the Space Telescope Science
Institute, the National Aeronautics and Space Administration under Grant
No. NNX08AR22G issued through the Planetary Science Division of the NASA
Science Mission Directorate, the National Science Foundation under Grant
No. AST-1238877, the University of Maryland, and Eotvos Lorand
University (ELTE) and the Los Alamos National Laboratory.

We also acknowledge support from Science and Technology Facilities Council
Consolidated Grant ST/P 000541/1.

Finally, we thank an anonymous referee for comments which have significantly
improved the quality of this paper.









\bibliographystyle{mnras} 
\bibliography{high_z_lf_v12} 
\bsp	
\label{lastpage}
\end{document}